\documentclass{nature}
\usepackage{amsmath,amssymb,graphicx}
\usepackage{soul}
\usepackage{caption}
\usepackage{graphicx}
\usepackage{color}

\makeatletter
\let\saved@includegraphics\includegraphics
\AtBeginDocument{\let\includegraphics\saved@includegraphics}
\renewenvironment*{figure}{\@float{figure}}{\end@float}
\makeatother

\title{Interlayer Exciton Laser with Extended Spatial Coherence in an Atomically-Thin Heterostructure}

\author{Eunice Y. Paik,$^{1*}$ Long Zhang,$^{1}$ G. William Burg,$^{2}$\\ Rahul Gogna,$^{3}$ Emanuel Tutuc,$^{2}$ Hui Deng,$^{1*}$}

\begin{document}

\maketitle

\begin{affiliations}
 \item Department of Physics, University of Michigan, Ann Arbor, Michigan 48109-2122
 \item Microelectronics Research Center, Department of Electrical and Computer Engineering, Austin, TX 78758
 \item Applied Physics Program, University of Michigan, Ann Arbor, Michigan 48109-1040
\end{affiliations}

\begin{abstract}
Two-dimensional semiconductors have emerged as a new class of materials for nanophotonics for their strong exciton-photon interaction and flexibility for engineering and integration. Taking advantage of these properties, we engineer an efficient lasing medium based on dipolar interlayer excitons, in rotationally aligned atomically thin heterostructures. Lasing is measured from a  transition metal dichalcogenide hetero-bilayer integrated in a silicon nitride grating resonator. A sharp increase in the spatial coherence of the emission was observed across the lasing threshold. The work establishes interlayer excitons in two-dimensional heterostructures as a silicon-compatible coherent medium. With electrically tunable light-matter interaction strength and long-range dipolar interactions, these interlayer excitons promise both applications to low-power, ultrafast laser and modulators and rich many-body quantum phenomena.
\end{abstract}

Semiconductor lasers are ubiquitous in today's technology because they are compact, cover a wide range of wavelengths, and allow efficient electrical pumping and fast electrical modulation\cite{Calvez_Semiconductor_2012}.
They are predominantly based on traditional III-V quantum wells. To achieve lower power consumption, more compact size, and a higher degree of integration on the silicon platform, there has been tremendous effort to develop alternative gain materials and structures, such as nanowire lasers\cite{Eaton_Semiconductor_2016}, spasers\cite{Noginov_Demonstration_2009}, and photonic crystal lasers\cite{Noda_Seeking_2006}. However, tunability, electrical pumping and silicon-chip integration remain common challenges for all these approaches.

Recently, monolayer transition metal dichalcogenide crystals (TMDCs) have emerged as a new class of materials for semiconductor lasers, for they are atomically thin, feature strong exciton emission \cite{Mak_Atomically_2010,Splendiani_Emerging_2010} and can be integrated with diverse materials, including silicon compounds \cite{Geim_Van_2013}.
% their many unique properties. Electrons in TMDs are two-dimensional in nature. Therefore, atomically-thick heterostructures can be created layer-by-layer, integrated with diverse substrates\cite{Geim_Van_2013}, and readily doped for electrical control\cite{Jones_Optical_2013,Sidler_Fermi_2016}, allowing unprecedented flexibility in engineering and integration.
%Excitons at the K-valleys are tightly bound with binding energies on the same order of the bandgap\cite{He_Tightly_2014,Chernikov_Exciton_2014}, promising room temperature exciton devices with high optical gain\cite{Chernikov_Population_2015,Su_How_2008,Fogler_High-temperature_2014,Liu_Strong_2014}.
%Spin-valley locking of the electrons promise novel valleytronic applications\cite{Mak_Valley_2014,Zhang_Electrically_2014,MacNeill_Breaking_2015,Rivera_Valley-polarized_2016,Schaibley_Valleytronics_2016}.
Previous studies have assessed lasing in monolayer TMDCs from the non-linear intensity dependence and linewidth reductions as a function of pump power~\cite{Wu_Monolayer_2015,Ye_Monolayer_2015,Salehzadeh_Optically_2015,Li_Room-temperature_2017,Shang_Room-temperature_2017,Zhao_High-Temperature_2018}. However, the photon flux appears to be below the stimulated scattering threshold~\cite{Reeves_2D_2018}. The emission often saturates soon after reaching the linear regime of operation. Spatial coherence -- one of the quintessential characters of laser \cite{Mandel_Coherence_1965} -- has not been studied. Hence it is difficult to exclude localized excitons, such as point-defects, as the source of the observed nonlinear power dependence. %The very large exciton oscillator strength of monolayer TMDs enables high gain, but at the same time may make the system prone to amplified spontaneous emission, which competes for gain and introduces instability.
Moreover, with only a monolayer as the gain medium, tunability is limited and vertical p-n junctions are not possible without contacting with other doped semiconductors.

In contrast, heterostructures open the door to engineering of the band structures and exciton states. In particular, spatially indirect excitons in heterostructures have been intensively studied  \cite{Butov_Condensation_1994,Fang_Strong_2014} for they feature a static dipole with long-range diople interactions that may give rise to rich manybody quantum phenomena \cite{Fogler_High-temperature_2014}; both the dipole interaction and exciton-light interaction strengths are tunable by external electrical fields, important for excitonic circuit applications \cite{butov_excitonic_2017}.  However, the reduced oscillator strength of spatial indirect excitons typically render them dark and hard to access.
%Similar spatially indirect excitons have been formed in GaAs \cite{Butov_Condensation_1994} and TMDC \cite{Fang_Strong_2014} double quantum wells, which are dark excitons with a long lifetime and therefore have been intensively studied for applications to excitonic devices \cite{butov_excitonic_2017} and for many-body quantum phases \cite{Fogler_High-temperature_2014}.

Here, we show that in rotationally aligned two dimensional (2D) hetero-bilayers \cite{Zhang_Interlayer_2017} integrated on a SiN cavity (Fig.~1), interlayer excitons form a bright and efficient gain medium, supporting lasing with extended spatial coherence at a low band inversion density. %Our laser device consist of a WSe$_{2}$-MoSe$_{2}$ hetero-bilayer on a silicon nitride (SiN) grating cavity (Fig.~1). %Schematic and optical microscope images of the device are shown in Fig. 1A and 2A. Consequently,
As illustrated in Fig.~1B, with precise alignment of the energy bands and sub-nanometer separation of the two monolayers, the inter-layer excitons retain a sufficiently large oscillator strength. With type-II band alignment, the hetero-bilayer forms a three-level system that allows efficient pumping through the intra-layer exciton resonances followed by rapid electron transfer to a lower-energy, empty conduction band~\cite{Rigosi_Probing_2015} (Fig.~1C). As a result, population inversion is readily achieved at the reduced band gap while avoiding fast intra-layer radiative loss of the carriers. Moreover, unlike those used for monolayer exciton lasers, the cavity mode in our device fully covers the hetero-bilayer, allowing gain for the full area of the bilayer and supporting extended spatial coherence (Fig.~1A).  %, straightforward integration with the hetero-bilayer \cite{Zhang_Photonic-crystal_2018}, and deep sub-wavelength total thickness of the device.
We observe lasing accompanied by a sharp increase in the spatial coherence length as the photon occupancy exceeds unity. The emission intensity increases nonlinearly by over a hundred-fold across the threshold, and continues to increase linearly with the pump power up to the highest power used without saturation. Our results establish 2D inter-layer excitons in engineered hetero-bilayers as an efficient lasing medium, which, compared to monolayer excitons, feature an electrically tunable long-range dipole interaction and oscillator strength~\cite{Jones_Optical_2013,Rivera_Valley-polarized_2016}, much more robust valley polarization%~\textcolor{red}{[Add Long's paper once on arxiv]}
, and a type II band alignment well-suited for electrical injection via an atomically thin bilayer p-n junction~\cite{Lee_Atomicall_2014,Withers_Light-emitting_2015,Ross_Interlayer_2017}.

%\section{Results}
%\paragraph*{Results}
%%Sample structure
The lasing device is comprised of a rotationally aligned WSe$_{2}$-MoSe$_{2}$ hetero-bilayer placed on a silicon nitride (SiN) grating resonator, as illustrated in Fig. 1a.  The WSe$_{2}$-MoSe$_{2}$ hetero-bilayer has type II band alignment, where inter-layer excitons are the lowest energy electronic excitations (Fig.~1c).
%Similar spatially indirect excitons have been formed in GaAs \cite{Butov_Condensation_1994} and TMDC \cite{Fang_Strong_2014} double quantum wells, which are dark excitons with a long lifetime and therefore have been intensively studied for applications to excitonic devices \cite{butov_excitonic_2017} and for many-body quantum phases \cite{Fogler_High-temperature_2014}. In our hetero-bilayer, however, bright inter-layer excitons are formed.
To form bright interlayer excitons, we precisely align the crystal axis of the WSe$_{2}$ and MoSe$_{2}$ monolayers to within $1 ^{\circ}$ of relative rotation via a polymer transfer technique. In this way, the valleys of the two monolayers overlap in momentum space to form a direct band-gap (Fig. 1b). %The alignment accuracy has been confirmed with second harmonic generation (Fig. S1).
The alignment accuracy can be confirmed with second harmonic generation (Supplementary Fig. S1).
The electron and hole in the bilayer, although separated into two different mono-layers, are only less than a nanometer apart. Consequently, the interlayer exciton remains as a good emitter; its oscillator strength, reduced by two orders of magnitude from that of intralayer ones~\cite{Ross_Interlayer_2017}, is still comparable to that of excitons in III-V semiconductors.

At the same time, as illustrated in Fig. 1c, carriers can be injected into the hetero-bilayers efficiently via the intra-layer exciton resonance, followed by rapid electron transfer, on the order of 10-100 fs ~\cite{Rigosi_Probing_2015}, to the empty conduction band of MoSe$_2$. As a result, band inversion can be established at the smaller, interlayer bandgap. Once separated into the two mono-layers, radiative recombination as well as valley depolarization, are also strongly suppressed, rendering long inter-layer exciton lifetimes on the order of 1~ns~(Supplementary Fig. S2). Photoluminescence (PL) from the hetero-bilayer shows much stronger inter-layer exciton emission than the intra-layer emission (Fig.~2b), confirming a sufficiently large oscillator strength and inter-layer exciton population built-up. Spatially resolved PL shows uniform emission from the inter-layer (intra-layer) excitons in the bilayer (monolayer) regions (Supplementary Fig. S3). %To demonstrate spectral and spatial isolation of the interlayer emission, we performed a spatial mapping of the sample emission (Fig. 1d). It shows enhanced interlayer exciton emission and suppressed intralayer emission in the heterostructure region of the sample.

%%Sample structure: cavity
The grating cavity provides optical feedback when the excitons are coupled to its resonances.
The cavity modes are sensitive to both the electric field propagation and polarization directions. We define the propagation (polarization) direction along the bar as x (TE) and cross-bar as y (TM), as illustrated in Fig. 2a.
We tune the grating period $\Lambda$, thickness $h$ and fill factor $g$ to obtain a high quality factor for the TE mode, and match it to the exciton resonance at zero in-plane wavenumber $k=0$.  The hetero-bilayer lies directly on the grating where the evanescent field remains strong \cite{Zhang_Photonic-crystal_2018}.%\hui{add Rahul's paper once on arxiv}.
The TM cavity modes are far blue detuned from the excitons, therefore the TM excitons modes are not affected by the cavity (Supplementary Fig.~S4).

We confirm the TE-cavity modes by measuring the empty-cavity dispersion with angle-resolved reflectance spectroscopy, which agrees well with the simulation by the rigorous coupled wave analysis (RCWA), as shown in Fig.~2c. The TE mode quality factor from the simulation is around 3,000. However, the actual cavity Q-factor is presumably lower than the simulated value due to fabrication imperfections.
From the reflectance spectra of the empty cavity, we estimate a Q-factor between 500 and 680  (Fig.~2c inset), but the exact value is difficult to determine due to low contrast and white-light noise. The PL spectral linewidth from the device corresponds to a Q-factor around 630 (Fig.~2d). % and polymer residue resulting from the TMDC heterostructure transfer process.

%The TM mode is far blue-detuned from the interlayer exciton emission energy and thus does not exhibit lasing behavior. Therefore, we use TM mode to analyze the uncoupled interlayer exciton emission.

The hetero-bilayer allows efficient optical pumping through the intralayer exciton resonances, which are far above  the resonances of the interlayer excitons or the cavity. To establish lasing, we pump the device with a 1.7 eV pulsed Ti:Sapphire laser (80~MHz repetition rate, 150~fs pulse width) and first measure the power-dependent PL spectra at 5~K using angle-resolved micro-PL (Supplementary Fig. S5).

The PL from the cavity mode at $k\sim 0$ brightens up rapidly as the pump power increases. We show in Fig.~3a along-bar angle resolved PL above threshold, overlaid with the simulated cavity modes (crosses). The red-shifted lasing emission compared to the bare cavity mode is to be expected because transferring monolayers onto the grating cavity red-shifts the modes by several meV's.
%The PL from the cavity mode at $k\sim 0$ brightens up rapidly as we increase the pump power. We compare in Fig.~2a and 2b the along-bar angle resolved PL at two pump powers, overlaid with the simulated PCC modes (crosses). Fig.~2a corresponds to  0.55P$_{th}$, measured with the lowest pump power where such a spectrum can be obtained above the noise in our experiment, which shows weak emission spread out in $k$. In comparison, Fig. 2b is taken at 3.33P$_{th}$ and shows strong, sharp emission at $k\sim 0$.
%The laser emission at 2.67P$_{th}$ is centered at the $k=0$ PCC mode at 1.354~eV. It features a narrow linewidth (FWHM) of 1.652\pm ???~meV, determined by fitting a Lorentzian lineshape to the PL emission spectrum.
Integrating over $k_{x}$ = $\pm$ 0.48~$\mu$m$^{-1}$, we obtain the photon occupancy $I_p(k\sim 0)$ after counting for the independently measured collection efficiency of the optical path~\cite{photon_number}. The $I_p(k\sim 0)$  vs pump power shows clearly a super-linear dependence as $I_p(k\sim 0)$ approaches one, consistent with the onset of the stimulated scattering into the cavity mode (Fig. 3b).

The pump power at the threshold of $I_p(k\sim 0)=1$ is $P_{th}$= 0.18~$\mu$W. From $P_{th}$ and the typical absorption efficiency of 20\% by monolayer WSe$_2$, we obtain the threshold carrier density $ n_{th}= 5.7 \times 10^{10}~\mbox{cm}^{-2}$, in good agreement with the density required for the transparency condition $n_{tr} \sim 8 \times 10^{10} ~\mbox{cm}^{-2}$ (see supplementary information for calculations of $ n_{th}$ and $ n_{tr}$).
Far above threshold, the output intensity becomes linear with the pump power without saturation up to the highest power used for TE measurements at $P= 28P_{th}$. %The line-width plateaus, possibly due to strong dephasing in the gain medium and mode competition due to the lack of lateral confinement.

Accompanying the super-linear increase in the emission intensity at threshold, the linewidth of the emission drops sharply, as shown in Fig. 3b, signifying the increase of temporal coherence. Linewidth at lower excitation powers than $\sim$0.1~$\mu$W is presumably broader but our detectors are not sensitive enough to detect the emission. The saturation and slight increase of the linewidth with increasing power above threshold may be due to interactions among the carriers and spatial mode competition\cite{Spivak_Is_2007}.

In stark contrast with TE emission, the TM polarized emission is not coupled to the cavity mode and does not show a threshold behavior. The emission becomes detectable only at high pump powers. An example is shown in Fig.~3c for $P=10~\mu$W. The emission spreads uniformly in $k$ over the numerical aperture of our collection optics and has a broad linewidth of about 70~meV. With increasing pump power, the total integrated emission intensity increase sub-linearly with pump power (Fig. 3d). Under the same excitation power, the integrated intensities from TM and TE polarized emission are comparable, as shown in Fig. 3b and Fig. 3d. However, the TM emission is spread over a broad range in energy and $k$ distribution, while the TE emission is concentrated to energy and $k$ ranges one to two orders of magnitude smaller as a result of stimulated scattering.

%%Spatial coherence properties
To confirm extended coherence expected of a laser with a 2D gain medium, we study the first-order spatial coherence function $g^{(1)}(\textbf{r},-\textbf{r})$  defined as follows~\cite{Mandel_Coherence_1965}:
\begin{equation}\label{eq.g1}
g^{(1)}(\textbf{r}_{1},\textbf{r}_{2}) = \dfrac{G^{(1)}(\textbf{r}_{1},\textbf{r}_{2})}{\sqrt{G^{(1)}(\textbf{r}_{1},\textbf{r}_{1})G^{(1)}(\textbf{r}_{2},\textbf{r}_{2})}}
\end{equation}
$G^{(1)}$ is the first-order correlation function and corresponds to:
\begin{equation}
G^{(1)}(\textbf{r}_{1},\textbf{r}_{2}) = \mbox{tr}\{\rho E^{(-)}(\textbf{r}_{1})E^{(+)}(\textbf{r}_{2})\}
\end{equation}
where $E^{(+)}$ and $E^{(-)}$ are field creation and annihilation operators, respectively.
%Previous studies
Although spatial coherence properties have been extensively studied in semiconductor photon lasers, exciton-polariton lasers~\cite{Deng_Spatial_2007} and plasmon lasers~\cite{Hoang_Millimeter-Scale_2017}, coherence of TMD lasers have not been studied thus far, making it difficult to rule out localized excitons as a source of lasing. The large spatial area of the grating resonator and large photon flux above threshold allow us to investigate the spatial coherence of the interlayer exciton emission.

First order spatial coherence measurements were performed using a continuous wave excitation laser and the retro-reflector Michelson interferometer setup~\cite{Daskalakis_Spatial_2015}. After filtering out the scattered laser, emission from the lasing device was sent to the interferometer which interferes an image of the sample with a centro-symmetrically inverted version (Fig. 4a inset).
Because of a small angle difference between the two beams,  interference fringes are formed (Fig. 4a) corresponding to slightly varying path length difference $\frac{2\pi}{\lambda_{0}}z_0(\textbf{r})$ at different positions $\textbf{r}$ across the images; $z_0$ is the initial position of the retro-reflector that produces the inverted image. The interferometer output is recorded by a charge-coupled camera with intensities described by:
\begin{align}
\label{eq.interference}
I^{int}(\textbf{r}) = I(\textbf{r}) + I(\textbf{-r}) + 2\sqrt{I(\textbf{r})I(\textbf{-r})} g^{(1)}(\textbf{r},-\textbf{r}) \sin\left(\frac{2\pi}{\lambda_{0}}\left(z - z_{0}\right)\right).
\end{align}
Here $I(\textbf{r})$ and $I(\textbf{-r})$ are intensities from the mirror and retro-reflector arms, respectively, and are measured by blocking one of the arms of the interferometer. $z$ is the position of the retro-reflector. $g^{(1)}(\textbf{r},-\textbf{r})$ is the first order spatial coherence for two positions separated by $2r$, and is proportional to the visibility of the interference fringe. To obtain the visibility or $g^{(1)}(\textbf{r},-\textbf{r})$, we scan the position $z$ of the retro-reflector and record the sinusoidal oscillation of $I^{int}(\textbf{r})$ vs. $z$ at each $\textbf{r}$, as shown in Fig. 4b. Fitting the oscillation at each $\textbf{r}$,  %Visibility can be extracted by fitting the modulation with a sinusoidal function and determining the $I_{\mbox{max}}$ and $I_{\mbox{min}}$ of the fringe pattern.
%Visibility of the interference fringes ($V(\textbf{r})$) is related to $g^{(1)}(\textbf{r},-\textbf{r})$ in the following way:
%\begin{equation}\label{eq.vis}
%V(\textbf{r}) = \frac{I_{\mbox{max}}-I_{\mbox{min}}}{I_{\mbox{max}}+I_{\mbox{min}}} = \frac{2\sqrt{I\left(\textbf{r}\right)I\left(-\textbf{r}\right)}}{I\left(\textbf{r}\right)+I\left(-\textbf{r}\right)}g^{(1)}(\textbf{r},-\textbf{r})
%\end{equation}
%By finding the visibility of every pixel on the interference image like the one shown in Fig. 3a,
we construct a map of spatial coherence, $g^{(1)}(\textbf{r},-\textbf{r})$. Below threshold, the emission is too weak for $g^{(1)}$ measurements. The top panels of Fig. 4c and 4d show examples of spatial coherence maps near and above threshold, respectively. Near threshold, the map is rather noisy without a clear patten of $g^{(1)}(\textbf{r},-\textbf{r})$ vs. $\textbf{r}$. Above threshold, a clear pattern emerges, showing a high $g^{(1)}(\textbf{r},-\textbf{r})$ near $r\sim 0$ that decays with increasing $r$ and extends above the background fluctuations to about the laser spot size.

%Note that $x$ is the along-bar direction and $y$ is the cross-bar direction, as shown in the inset of Figure~\ref{fig4}(a).
To study the functional dependence of $g^{(1)}$, we average over $y =~\pm 0.2~\mu$m and obtain $g^{(1)}(x,-x)$. The bottom panels of Fig. 4c and 4d show examples for near and above threshold, respectively. A clearly slower decay of $g^{(1)}(x,-x)$ with $x$ is measured above threshold. The $g^{(1)}(x,-x)$ vs. $x$ is fit well by a Gaussian function with the standard deviation $\sigma$ as a fitting parameter. From the fits, we obtain the coherence length $\lambda_{c} = 2\sqrt{2\pi}\sigma$. As seen in Fig. 4e, $\lambda_{c}$ increases sharply across the threshold, from $2.38~\mu$m near threshold to about $5~\mu$m above threshold,  confirming the formation of extended spatial coherence in the laser. % but at higher pump powers $\lambda_{c}$ plateaus around 4.87 $\mu$m as shown in Figure~\ref{fig5}.

Further above threshold, $\lambda_{c}$ remains about the same, possibly limited by the laser spot size; it decreases slightly at the highest powers, possibly because of competition of multiple spatial mode in the absence of lateral confinement potentials. We note that, the measured $g^{(1)}(x,-x)$ is much lower than the actual value due to difficulty in achieving good alignment. The emission from our ultra-compact device is necessarily weak and the detector efficiency at 1.36~eV is poor, hence it is difficulty to simultaneously achieve good time and spatial overlap of the interference signal with the asymmetric interferometer. With a fully symmetric interferometer without inverting one of the images, we obtain visibility for $g^{(1)}(\tau = 0)$ close to 0.8 (supplementary Fig. S6). %Under pulsed excitations, where the time overlap can be aligned precisely independently, we are able to measure $g^{(1)}(x=0)$ up to ???.
%In comparison, below threshold, $\lambda_{c}$ is limited by the spatial extent of the cavity mode, which can be estimated by the cavity linewidth and the corresponding $\Delta k$. Using the linewidth of 2.15~meV measured at the lowest pump power and the along-bar cavity dispersion, the estimated uncertainty in $k_{x}$ is $\Delta k_{x} \sim$1.2 $\mu$m$^{-1}$, which gives the spatial extent of $\Delta x \sim $0.4 $\mu$m.

In conclusion, we demonstrate a two-dimensional WSe$_{2}$-MoSe$_{2}$ hetero-bilayer laser on a grating cavity. Nonlinear increase in the output intensity and narrowing of the emission linewidth are measured as the photon number in the lasing mode reaches the order of unity. The carrier density at threshold matches that for transparency condition and is reduced compared to in monolayers due to the reduced interlayer band gap. The full coverage of gain medium by the cavity field enabled a relatively large photon flux above threshold and a measurement of the spatial coherence of the device. A sharp increase of the spatial coherence to around 5~$\mu$m above threshold was measured, confirming the lasing emission originates from an extended 2D gain medium -- the inter-layer excitons.

The hetero-bilayer laser can be improved by reducing the inhomogeneous broadening of the gain medium with hexagonal boron-nitride encapsulation%\cite{Cadiz_Excitonic_2017}
, improving the cavity Q, and reducing mode competition with lateral confinement of the cavity modes
%\cite{Vuckovic_Design_2001,Yokouchi_Two-dimensional_2003}
. Using cavities with lateral rotational invariance would allow valleytronic applications of the inter-layer excitons laser. Different combinations of van der Waals materials can be used to create inter-layer exciton lasers of different wavelengths%\cite{Kozawa_Evidence_2016,Okada_Direct_2018,Kunstmann_Momentum-space_2018}
. Electrical tuning of the oscillator strength may allow fast modulation of the laser and electrical injection can be implemented via atomically thin, bilayer p-n junctions \cite{Lee_Atomicall_2014,Withers_Light-emitting_2015,Ross_Interlayer_2017}.%\cite{Lee_Atomicall_2014,Withers_Light-emitting_2015,Ross_Interlayer_2017}.
Coherent indirect exciton gases may be explored by adiabatic electrical tuning \cite{shahnazaryan_adiabatic_2015}.

\begin{methods}
	
\subsection{Sample fabrication.}
To fabricate the grating cavity, we first grew a SiN film with a SiO$_{2}$ buffer on Si substrate using low-pressure chemical vapor deposition, then patterned it using electron beam lithography and created the grating bars by plasma dry etching. The grating parameters indicated in Fig. 1a are as follows: $d = 1475$ nm, $t = 113$ nm, $h = 60$ nm, $\lambda = 615$ nm and $g = 50$ nm.
The individual WSe$_{2}$ and MoSe$_{2}$ monolayers were mechanically exfoliated onto a SiO$_{2}$ substrate using PDMS. The exfoliated monolayers were stacked into a heterostructure using a high accuracy rotational alignment method~\cite{Kim_van_2016}. First, the MoSe$_{2}$ was picked up with a PDMS/PPC stamp under an optical microscope. Then, the crystal axes of MoSe$_{2}$ and WSe$_{2}$ were rotationally aligned to be 0 or 60 degrees prior to stacking. The stacked heterostructure was dropped down onto the PCC. Polymer residue was dissolved and the sample was annealed at 350$^{\circ}$C for a total of 7 hours.

\subsection{Hetero-bilayer twist angle.}
We can verify the twist angle of the heterobilayer aligned under the optical microscope by angle-dependent second harmonic generation (SHG) measurements. Supplementary Fig. S1 shows an optical microscope image of two different samples and the corresponding angle dependent SHG measurements (Supplementary Fig. S1). By fitting the SHG pattern with a cos$^{2}$(3$\theta$) function, where $\theta$ is the angle between the armchair direction of the monolayer and the polarization direction of the beam, we can obtain the twist angle. We did not measure the SHG for the  hetero-bilayer prior to putting it on the grating, but from experience, we found that the straight edges of exfoliated monolayers reliably correspond to the arm-chair axis of the crystal. Therefore, we aligned the two straight flake edges under the optical microscope.

\subsection{Optical measurements.}
We measure the lifetime of the TM emission using a Hamamatsu streak camera system. The sample was excited using a pulsed Ti:sapphire laser. The emission was polarization selected for the TM direction and sent to the streak camera. As shown in supplementary Fig. S2, a linecut of the streak camera spectrum was fitted with a bi-exponential function to determine the lifetime. The fitted lifetime is around 2 ns.
Spatially resolved PL mapping of the hetero-bilayer lasing device was shown in supplementary Fig. S3. The sample was excited with a 633 nm continuous wave (CW) laser and selected for TE polarization. With an Attocube ANC 300 piezo stage, we scanned the laser spot around various regions of the sample. We applied spectral band-pass filter to the PL spectra to select out the emission from bi-layer, WSe$_{2}$, and MoSe$_{2}$ regions.
Supplementary Fig. S5 shows the schematic of the optical setup used for the angle-resolved PL/reflection and coherence measurements. The sample was cooled to 5 K using a Montana Instruments Fusion 2 cryostat. Fourier-space imaging was used to measure angle-resolved reflection and micro-PL of the device. For reflection, a tungsten halogen lamp was used. For micro-PL a pulsed Ti:sapphire laser near-resonant with the WSe$_{2}$ A-exciton (1.7 eV) was used to excite the sample. The emission was collected using a 0.55 NA objective lens, passed through a long-pass filter to filter out the excitation laser and a linear polarizer to selectively measure TE and TM modes, and sent to a Princeton Instruments spectrometer with a measured spectral resolution of 0.3 nm.
Spatial coherence measurement was performed using a retro-reflector Michelson Interferometer setup as shown in supplementary Fig. S5. Emission rid of scattered pump laser light was sent to a 50:50 beam-splitter which divided the light into two paths, the mirror path and the retro-reflector path. In order to change the time difference ($\tau$) between the two paths, the retro-reflector is mounted on a stepper motor which can have a step size as small as 50 nm (about 0.167 fs). The interference pattern is collected by a Princeton Instruments eXcelon CCD camera.

\end{methods}

%% Put the bibliography here, most people will use BiBTeX in
%% which case the environment below should be replaced with
%% the \bibliography{} command.

% \begin{thebibliography}{1}
% \bibitem{dummy} Articles are restricted to 50 references, Letters
% to 30.
% \bibitem{dummyb} No compound references -- only one source per
% reference.
% \end{thebibliography}
\section*{References}
\bibliographystyle{naturemag}
\bibliography{paik_arxiv}

\begin{thebibliography}{10}
\bibitem{photon_number}
\bibinfo{note}{Photon occupancy per pulse ($I_p(k\sim 0)$) was estimated from
  the total count rate on the detector, $n_{c}$. The two values are related by:
  $I_{c} = \eta \rho f n_p(k\sim 0)$. Here $\eta \sim 10^{-7}$ is the total
  detection efficiency of the setup, which is independently calibrated by
  replacing the sample with a laser-coupled single mode fiber. $\rho \sim 1$ is
  the number of k-space modes within the integrated region. $f = 80$~MHz is the
  repetition rate of the pump laser.}
\end{thebibliography}

%% Here is the endmatter stuff: Supplementary Info, etc.
%% Use \item's to separate, default label is "Acknowledgements"

\begin{addendum}
 \item[Data availability] Data are available on request from the authors.
 \item[Acknowledgements] The authors acknowledge the support by the Army Research Office under Awards W911NF-17-1-0312.
 \item[Author contributions] H.D. and E.T. supervised the project. L.Z. fabricated the SiN grating. E.Y.P. and L.Z. exfoliated the WSe$_{2}$ and MoSe$_{2}$ monolayers. G.W.B made the rotationally aligned structure. R.G. designed the SiN grating device. E.Y.P. performed the measurements. E.Y.P. and H.D. performed data analysis and wrote the manuscript.
 \item[Competing interests] The authors declare that they have no
competing financial interests.
 \item[Correspondence] Correspondence and requests for materials
should be addressed to E.Y.P. and H.D.~(email: eypaik@umich.edu, dengh@umich.edu).
\end{addendum}

\pagebreak
%%
%% TABLES
%%
%% If there are any tables, put them here.
%%

\begin{figure}
	\centering
	\includegraphics[width=0.7\textwidth]{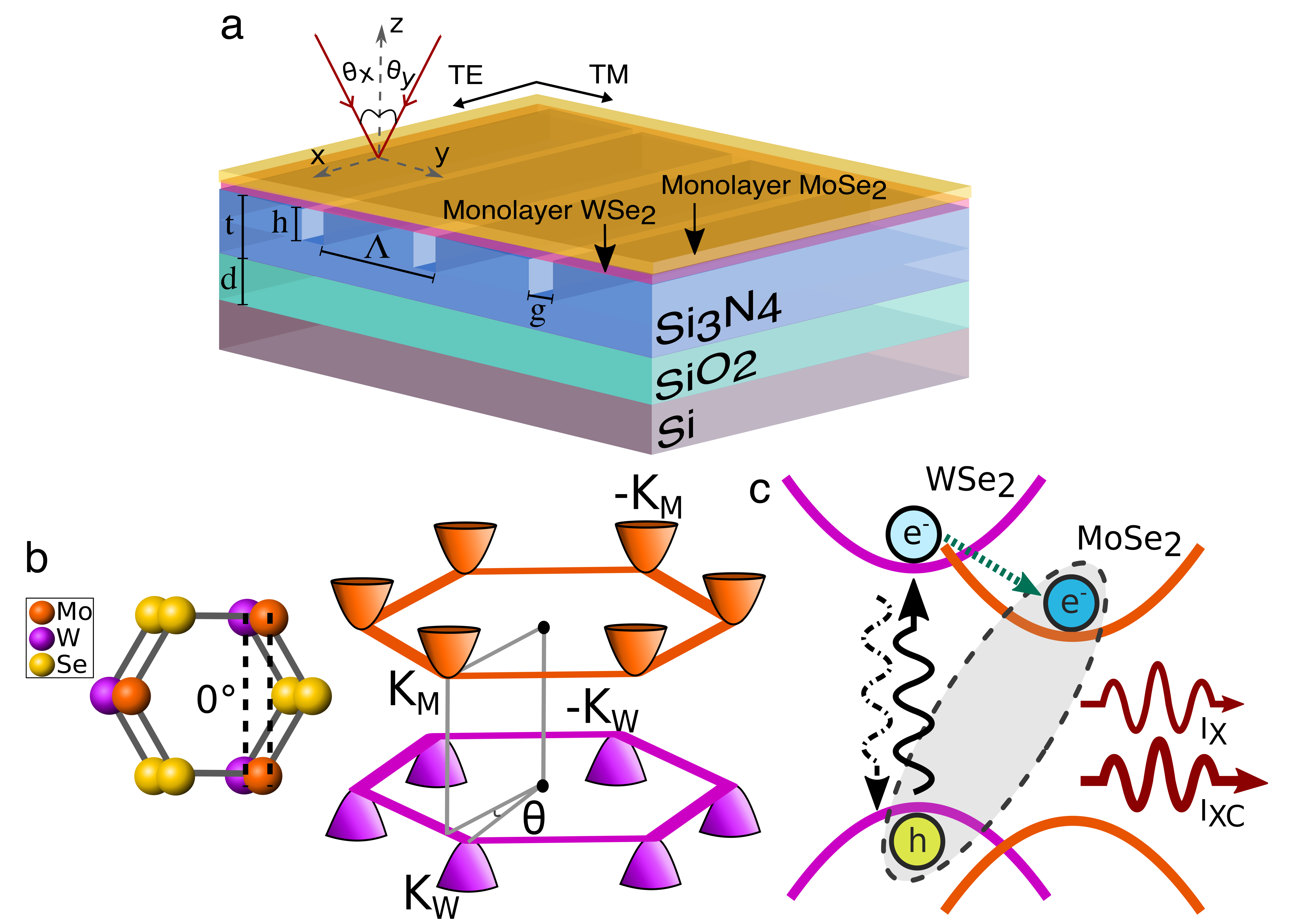}
	\caption{\textbf{Heterobilayer/grating cavity system.} \textbf{a}, Schematic of the laser device consisting of a hetero-bilayer on a grating cavity. The along-bar (cross-bar) direction and polarization are defined as $x$ ($y$) and TE (TM) respectively. Grating cavity design parameters are the following: total SiN thickness ($t$), SiO$_{2}$ thickness ($d$), grating thickness ($h$), grating period ($\Lambda$), gap width ($g$). \textbf{b}, Illustration of the rotationally aligned heterobilayer with twist angle $\theta =~ $0$^{\circ}$ (left), and the corresponding band structure with extrama at the K-valleys (right). Depending on the twist angle $\theta$, the hetero-bilayer has a direct or indirect bandgap. \textbf{c}, Band alignment and carrier dynamics of the hetero-bilayer. The hetero-bilayer has a type-II band alignment, forming a three level system for the injected carriers.  Intra-layer excitons are excited by a pump laser in the WSe$_{2}$ layer (solid wavy line). Some electrons transfer to the lower MoSe$_{2}$ conduction band on a fast (10-100 fs) time-scale (dotted line), while others recombine as intra-layer excitons with lifetimes around 1-10 ps (dash-dotted line). Without the cavity, the inter-layer excitons (dashed line) recombine with a lifetime on the order of 1 ns (I$_{\mbox{X}}$), and, with cavity enhancement, on the order of 1-10 ps (I$_{\mbox{XC}}$).}
\end{figure}

\begin{figure}
	\centering
	\includegraphics[width=0.65\textwidth]{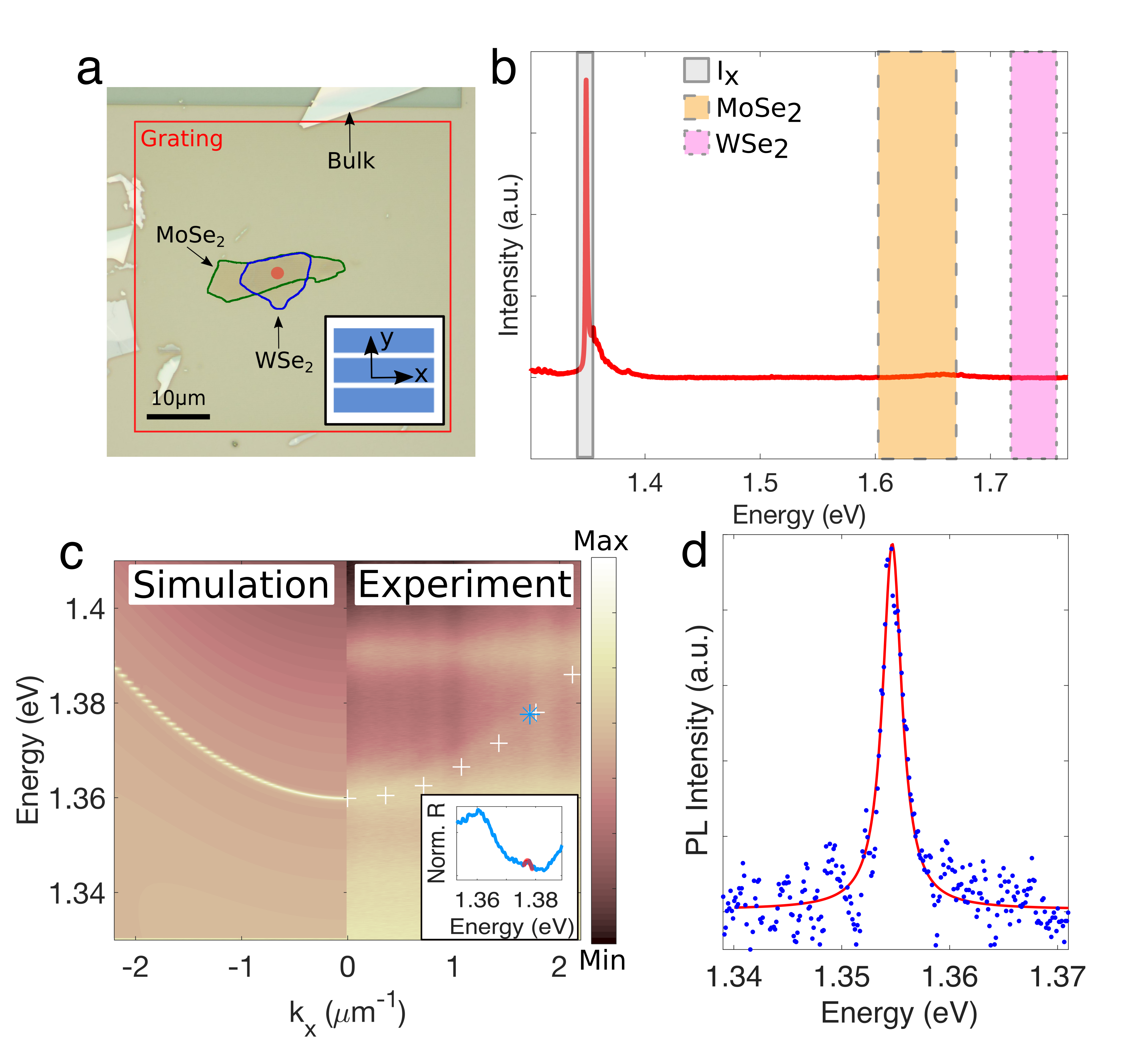}
	\caption{\textbf{Characterization of the heterobilayer/grating cavity system.} \textbf{a}, An optical microscope image of the WSe$_{2}$/MoSe$_{2}$ hetero-bilayer integrated on a grating cavity. The red square outlines the grating region and the red circle indicates approximately the laser spot size. Inset: direction of the grating bars. \textbf{b}, PL spectrum from the hetero-bilayer. The shaded boxes highlight the approximate spectral range of inter-layer, MoSe$_{2}$ and WSe$_{2}$ exciton emission. \textbf{c}, TE-polarized along-bar, angle-resolved, simulated (left) and measured (right) reflectance spectrum. The cavity mode dispersion is evident from the peaks in the reflectance spectrum. The overlaid crosses indicate the cavity mode from simulation. Inset: linecut of the spectrum around $k_{x} \sim 1.7~\mu$m$^{-1}$. Red line is a fit to the cavity mode. Star symbol marks the peak of the fitted cavity mode. \textbf{d}, PL spectrum near $k_{x} \sim 0$ at a pump power of $0.1~\mu$W. Red line is the Lorentzian fitting to the spectrum with a fitted linewidth of 2.2~meV.}
	\label{fig2}
\end{figure}

\begin{figure}	
\centering
	\includegraphics[width=0.7\textwidth]{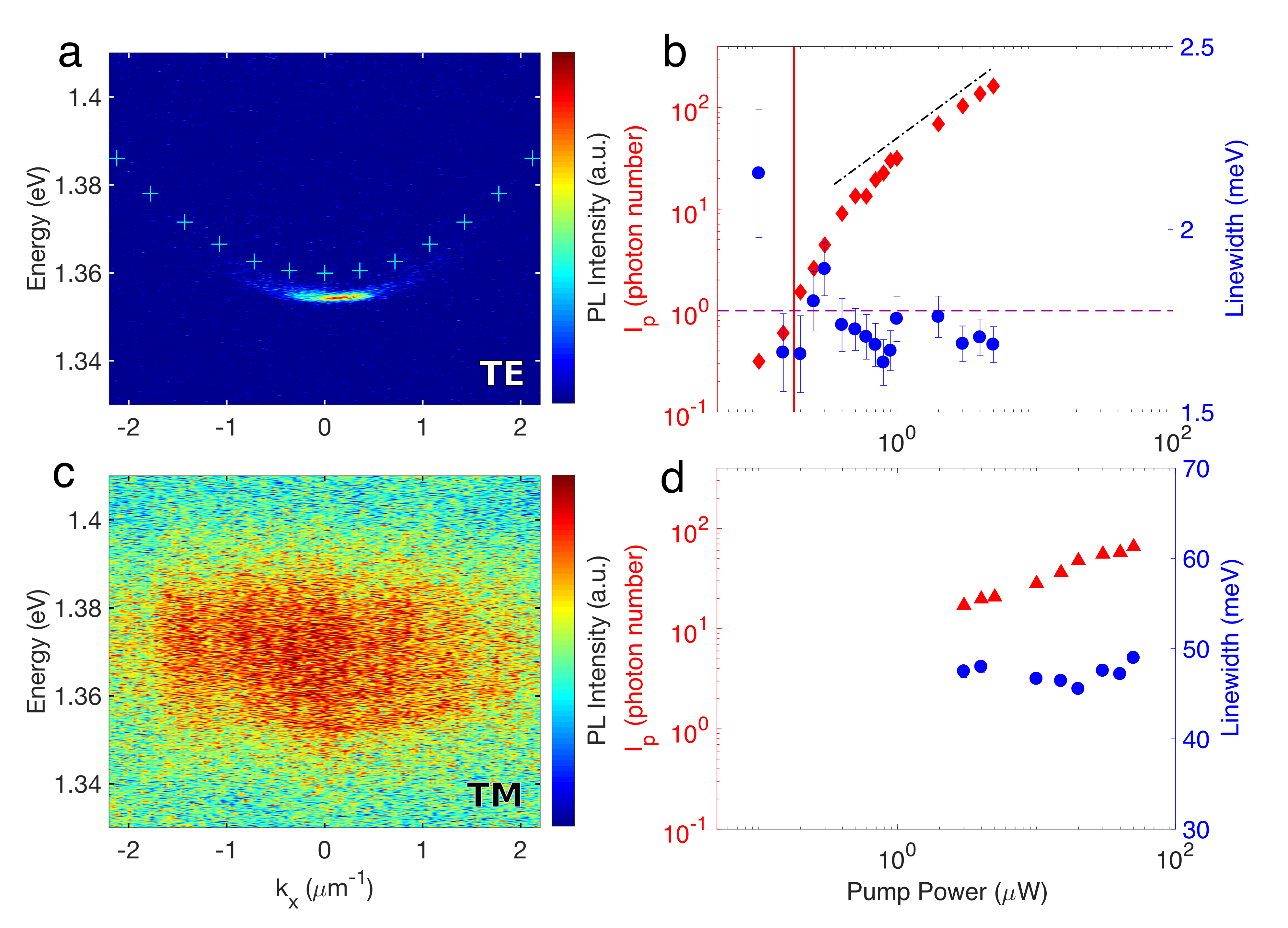}
	\caption{\textbf{Properties of the integrated hetero-bilayer/grating cavity system.} \textbf{a}, Angle resolved micro-PL spectra for the along-bar TE direction at 3.33P$_{th}$ with overlaid simulated empty cavity dispersion (crosses). The cavity resonances is red-shifted when the heterobilayer is placed on top. \textbf{b}, The intensity (red) and linewidth (blue) of the TE emission from the heterobilayer versus input pump power. The emission intensity is integrated over $|k_{x}|< 0.48~\mu$m$^{-1}$ and $|k_{y}|< 0.13~\mu$m$^{-1}$. The dot-dashed line indicates a linear line, the vertical red line marks P$_{th}$, and the horizontal purple line indicates $I_{p} = 1$. \textbf{c}, Angle resolved micro-PL spectra for the along-bar TM direction at P=10~$\mu$W. \textbf{d}, The pump power dependence of the TM emission intensities, integrated over $k_{x} \pm 2~\mu$m$^{-1}$ (red), and the TM emission line-width (blue).}
	\label{fig3}
\end{figure}

\begin{figure}
	\centering
	\includegraphics[width=1\textwidth]{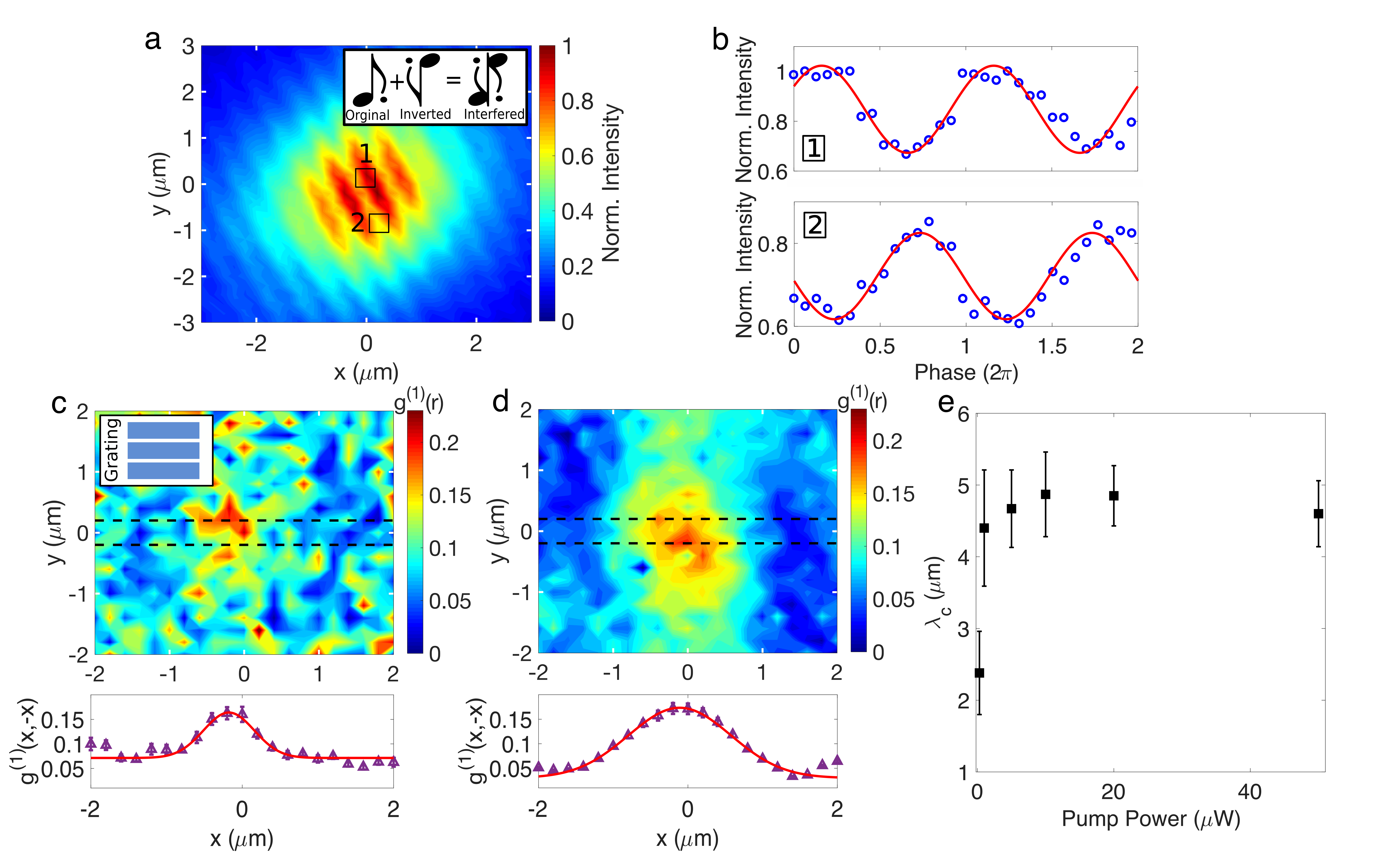}
	\caption{\textbf{First order coherence of the inter-layer exciton laser.} \textbf{a}, Typical interference pattern above P$_{th}$. Inset: illustration of centro-symmetrically interfered images. \textbf{b}, Intensity plots of single pixels (labeled as squares 1 and 2 in (a)) as the retro-reflector position is scanned over a phase of 4$\pi$. \textbf{c,d}, Top: maps of $g^{(1)}(\textbf{r},-\textbf{r})$ \textbf{c} near P$_{th}$ and \textbf{d} above P$_{th}$. Bottom: horizontal linecuts of $g^{(1)}(\textbf{r},-\textbf{r})$  integrated between the dashed lines. The red line is the Gaussian fit used to extract the coherence length $\lambda_{c}$. Inset: illustration of the grating bar direction. \textbf{e}, The coherence length $\lambda_{c}$ versus the pump power.}
	\label{fig4}
\end{figure}

\end{document}